\documentclass[11pt,referee]{aa}
\usepackage{graphics}

\def\beq{\begin{equation}}
\def\eeq{\end{equation}}
\def\bey{\begin{eqnarray}}
\def\eey{\end{eqnarray}}
\def\beqarray{\begin{eqnarray}}
\def\eeqarray{\end{eqnarray}}

\def\kms{\,{\rm {km\, s^{-1}}}}

\def\v200{V_{200}}

\begin{document}

   \thesaurus{( 11.01.2;       
                11.06.1;       
                11.05.2)}      

\input epsf
\title
{Star Formation and Chemical Evolution of Lyman-Break Galaxies}

\author{Chenggang Shu}
\offprints{C. Shu, cgshu@center.shao.ac.cn}
 \institute{
      1. Shanghai Astronomical Observatory, Chinese Academy of
Sciences, Shanghai 200030, P. R. China\\
      2. Max-Planck-Institut f\"ur Astrophysik
      Karl-Schwarzschild-Strasse 1, 85748 Garching, Germany\\
      3. National Astronomical Observatories, Chinese Academy of Sciences, P. R. China\\
      4. Joint Lab of Optical Astronomy, Chinese Academy of Sciences, P. R. China}

\date{Accepted ........
      Received ........}

\titlerunning{Star Formation and Chemical Evolution of LBGs}

\maketitle
\begin{abstract}
The number density and clustering properties of Lyman-break
galaxies (LBGs) observed at redshift $z\sim 3$ are best explained
by assuming that they are associated with the most massive haloes
at $z\sim 3$ predicted in hierarchical models of structure
formation. In this paper we study, under the same assumption, how
star formation and chemical enrichment may have proceeded in the
LBG population. A consistent model, in which the amount of cold
gas available for star formation must be regulated, is suggested.
It is found that gas cooling in dark haloes provides a natural
regulation process. In this model, the star formation rate in an
LBG host halo is roughly constant over about 1 Gyr. The predicted
star formation rates and effective radii are consistent with
observations. The metallicity of the gas associated with an LBG is
roughly equal to the chemical yield, or about the order of $1
Z_{\odot}$ for a Salpeter IMF. The contribution to the total
metals of LBGs is roughly consistent with that obtained from the
observed cosmic star formation history. The model predicts a
marked radial metallicity gradient in a galaxy, with the gas in
the outer region having much lower metallicity. As a result, the
metallicities for the damped Lyman-alpha absorption systems
expected from the LBG population are low. Since LBG halos are
filled with hot gas in this model, their contributions to the soft
X-ray background and to the UV ionization background are
calculated and discussed.
\end{abstract}

\begin{keywords}
galaxies: LBGs - galaxies: formation - galaxies: star formation -
galaxies: chemical evolution
\end{keywords}

\section {Introduction}

The Lyman-break technique (e.g. Steidel, Pettini \& Hamilton 1995)
has now been proved very successful in finding large numbers of
star forming galaxies at redshift $z\sim 3$ (e.g. Steidel et al.
1996, 1999b). The observed number density and clustering
properties of Lyman-break galaxies (hereafter LBGs, Steidel et al.
1998; Giavalisco et al. 1998; Adelberger et al. 1998) are best
explained by assuming that they are associated with the most
massive haloes at $z\sim 3$ predicted in hierarchical models of
structure formation (Mo \& Fukugita 1996; Baugh, Cole \& Frenk
1998; Mo, Mao \& White 1998b; Coles, et al. 1998; Governato, et
al. 1998; Jing 1998; Jing \& Suto 1998; Katz, et al.  1998;
Kauffmann, et al. 1998; Moscardini, et al. 1998; Peacock, et al.
1998; Wechsler, et al. 1998). This assumption provides a framework
for predicting a variety of other observations for the LBG
population. Steidel et al. (1999b and references therein) gave a
good summary of recent studies on this population including the
luminosity functions, luminosity densities, color distribution,
star formation rates, clustering properties, and the differential
evolution.

Assuming that LBGs form when gas in dark haloes settles into
rotationally supported discs or, in the case where the angular
momentum of the gas is small, settles at the self-gravitating
radius, Mo, Mao \& White (1998b) predict sizes, kinematics and
star formation rates and halo masses for LBGs, and find that the
model predictions are consistent with the current (rather limited)
observational data; Steidel et al. (1999a) suggest that the total
integrated UV luminosity densities of LBGs are quite similar
between redshift 3 and 4 although the slope of their luminosity
function might have a large change in the faint-end.

Furthermore, Steidel et al. (1999b) suggest that a ``typical" LBG
have a star formation rate of about $65h_{50}^{-2}\rm
{M_{\odot}yr^{-1}}$ for $\Omega_{0}=1$ and the star formation time
scale be the order of 1Gyr based on their values of E(B-V) as
pointed out by Pettini et al. (1997b) after adopting the reddening
law of Calzetti (1997). Recently, Friaca \& Terlevich (1999) use
their chemodynamical model to propose that an early stage (the
first Gyr) of intense star formation in the evolution of massive
spheroids could be identified as LBGs.

However, Sawicki \& Yee (1998) argued that LBGs could be very
young stellar populations with the age less than 0.2Gyr based on
the broadband optical and IR spectral energy distributions. This
is also supported by the work of Ouchi \& Yamada (1999) based on
the expected sub-mm emission and dust properties. It is worthy of
noting that the assumptions about the intrinsic LBG spectral shape
and the reddening curve play important roles in these results.

In this paper, we study how star formation and chemical enrichment
may have proceeded in the LBG population. As we will demonstrate
in Section 2, the observed  star formation rate at $z\sim 3$
requires a self-regulating process to keep the gas supply for a
sufficiently long time. We will show (in Section 2) that such a
process can be achieved by the balance between the energy feedback
from star formation and gas cooling. Model predictions for the LBG
population and further discussions about the results are presented
in Section 3, a brief summary is given in Section 4.

As an illustration, we show theoretical results for a CDM model
with cosmological density parameter $\Omega_{0}=0.3$, cosmological
constant $\Omega_\Lambda=0.7$. The power spectrum is assumed to be
that given in Bardeen et al. (1986), with shape parameter
$\Gamma=0.2$ and with normalization $\sigma_{8}=1.0$. We denote
the mass fraction in baryons by $f_{\rm B}=\Omega_{\rm
B}/\Omega_0$, where $\Omega_{\rm B}$ is the cosmic baryonic
density parameter. According to the cosmic nucleosynthesis, the
currently favoured value of $\Omega_{\rm B}$ is $\Omega_{\rm
B}\sim 0.019 h^{-2}$ (Burles \& Tytler 1998), where $h$ is the
present Hubble constant in units of 100 $\rm kms^{-1}Mpc^{-1}$,
and so $f_{\rm B}\sim0.063 h^{-2}$. Whenever a numerical value of
$h$ is needed, we take $h=0.7$. At the same time, we define
parameter $t_\star$ as the time scale for star formation in the
LBG population throughout the paper.

\section {Models}
\subsection{Galaxy Formation}

  In this paper, we use the galaxy formation scenario described
in Mo, Mao \& White (1998a, hereafter MMWa) to model the LBG
population. In this scenario, central galaxies are assumed to form
in dark matter haloes when collapse of protogalactic gas is halted
either by its angular momentum, or by fragmentation as it becomes
self-gravitating (see Mo, Mao \& White 1998b, hereafter MMWb, for
details). As described in MMWb, the observed properties of LBGs
can be well reproduced if they are assumed to be the central
galaxies formed in the most massive haloes with relatively small
spins at $z\sim 3$. As in MMWb, we assume that gas in a dark halo
initially settles into a disk with exponential surface density
profile.

When the collapsing gas is arrested by its spin, the central gas
surface density and the scale length of an exponential disk are
\beq
 \Sigma_{0} \approx 380h {\rm M}_{\odot}{\rm
pc^{-2}}\left({m_{\rm d} \over 0.05}\right)\left({\lambda \over
0.05}\right)^{-2} \left({V_{c} \over {\rm 250kms}^{-1}}\right)
\left[{H(z) \over {H_{0}}}\right], \eeq and \beq R_{\rm d} \approx
8.8h^{-1}{\rm kpc}\left({{\lambda} \over
{0.05}}\right)\left({{V_{c}} \over {{\rm 250kms}^{-1}}}\right)
\left[{H(z) \over {H_{0}}}\right]^{-1}, \eeq
 where $m_{\rm d}$ is the fraction of halo mass that settles into
the disk, $V_{c}$ is the circular velocity of the halo, $\lambda$
is the dimensionless spin parameter, $H(z)$ is the Hubble constant
at redshift $z$ and $H_{0}$ is its present value (see MMWa for
details). Since $H(z)$ increases with $z$, for a given $V_c$ disks
are less massive and smaller but have a higher surface density at
higher redshift. When $\lambda$ is low and $m_{\rm d}$ is high,
the collapsing gas will become self-gravitating and fragment to
form stars before it settles into a rotationally supported disk.
In this case, we will take an effective spin $\lambda\propto
m_{\rm d}$ in calculating $\Sigma_0$ and $R_{\rm d}$.

We take the empirical law (Kennicutt 1998) of star formation rate
(SFR) to model the star formation in high-redshift disks which is
 \beq\label{SFR_law}
 \Sigma_{\rm SFR} = a
\left({{\Sigma_{\rm gas}} \over {{\rm M}_{\odot}{\rm
pc^{-2}}}}\right)^{b} {\rm M}_{\odot}{\rm yr^{-1}}{\rm pc^{-2}} ,
\eeq
 where
  \beq a = {2.5 \times 10^{-10}}, ~~~ b=1.4 \eeq
 respectively. Here $\Sigma_{\rm SFR}$ is the SFR per unit area and
$\Sigma_{\rm gas}$ is the gas surface density. Note that this star
formation law was derived by averaging the star formation rate and
cold gas density over large areas on spiral disks and over
starburst regions (Kennicutt 1998). We will apply this law
differentially on a disk and also take into account the Toomre
instability criterion of star formation (Toomre 1964; see also
Binney \& Tremaine 1987).

For a given cosmogonic model, the mass function for dark matter
haloes at redshift $z$ can be estimated from the Press-Schechter
formalism (Press \& Schechter 1974): \beq {\rm d}N = -\sqrt{2
\over \pi}{\rho_{0} \over M}{\delta_{c}(z) \over \Delta(R)} {{\rm
d}\ln\Delta(R) \over {\rm d}\ln M}{\exp}\left[-{\delta_{c}^{2}(z)
\over {2\Delta^{2}(R)}}\right]{{\rm d} M \over M}, \eeq where
$\delta_{c}(z)=\delta_{c}(0)(1+z)g(0)/g(z)$ with $g(z)$ being the
linear growth factor at $z$ and $\delta_{c}(0)\approx 1.686$,
$\Delta(R)$ is the linear $rms$ mass fluctuation in top-hat
windows of radius $R$ which is related to the halo mass $M$ by
$M=(4\pi/3){\overline \rho}_{0}R^{3}$, with ${\overline\rho}_{0}$
being the mean mass density of the universe at $z=0$. The halo
mass $M$ is related to halo circular velocity $V_c$ by
$M=V_c^3/[10GH(z)]$. A detailed description of the PS formalism
and the related cosmogonic issues can be found in the Appendix of
MMWa.

From the Press-Schechter formalism and the $\lambda$-distribution
which is a log-normal function with mean ${\overline
{\ln\lambda}}=\ln 0.05$ and dispersion $\sigma_{\rm ln
\lambda}=0.5$ (see equation [15] in MMWa), we can generate Monte
Carlo samples of the halo distributions in the $V_c$-$\lambda$
plane at a given redshift and, using the star formation law
outlined above, assign a star formation rate to each halo. As in
MMWb, we select LBGs as the galaxies with the highest star
formation rate, so that the comoving number density for LBGs is
equal to the observed value, $N_{\rm LBG}=2.4 \times
10^{-3}h^{3}{\rm Mpc^{-3}}$ for the assumed cosmology at $z=3$, as
given in Adelberger et al. (1998). Here it is worth noting that
the model selection of LBGs we adopted is without the dust
extinction being considered. This implies that the contribution of
the dust is assumed to be uniform. But in fact, it could be very
different from galaxies to galaxies. So, our selection of LBGs may
not have one-to-one correspondence with the observed LBGs (Baugh
et al. 1999), but the selection should be correct on average.

\subsection{Cooling-Regulated Star Formation}

What regulates the amount of star-forming gas in a dark halo? In
the standard hierarchical scenario of galaxy formation (e.g. White
\& Rees 1978; White \& Frenk 1991, hereafter WF), gas in a dark
matter halo is assumed to be shock heated to the virial
temperature,
 \beq T=2.24 \times 10^6{\rm K} \left({V_c\over
250{\rm km\,s^{-1}}}\right)^2, \eeq
 as the halo collapses and
virializes. The hot gas then cools and settles into the halo
centre to form stars. As suggested in WF, the amount of cold gas
available for star formation in a dark halo is either limited by
gas infall or by gas cooling, depending on the mass of the halo.
For the massive haloes ($V_c\ga 200 {\rm km\,s^{-1}}$) we are
interested here, gas cooling rate is smaller than gas-infall rate,
and the supply of star-forming gas is limited by gas cooling (see
WF for details). It is therefore likely that gas cooling is the
main process that constantly regulates the SFR in LBGs.

To have a quantitative assessment, let us compare different rates
involved in the problem. Using equations (1)-(4) we can write the
SFR as
\begin{eqnarray}\label{MSFR}
\dot M_{\star }= {2\pi a\Sigma_{0}^{b}R_{\rm d}^{2} \over b^{2}}
\approx 2.33\times 10^{2}h^{-0.6}\left({m_{\rm d}\over
0.05}\right)^{1.4} \left({\lambda \over
0.05}\right)^{-0.8}\left({V_{c}\over {{\rm 250
km\,s^{-1}}}}\right)^{3.4}\left[{H(z) \over {H_{0}}}\right]^{-0.6}
{\rm M_{\odot}\,yr ^{-1}},
\end{eqnarray}
where $m_{\rm d}$ is the current gas content of the disk. The rate
at which gas is consumed by star formation is therefore
 \beq
  {\dot M}_{\rm SFR}=(1-R_{\rm r}) {\dot M}_\star,
   \eeq
    where $R_{\rm r}$
is the returned fraction of stellar mass into the ISM; we take
$R_{\rm r}=0.3$ for a Salpeter IMF (e.g. Madau et al. 1998).
According to WF, the heating rate due to supernova explosions
under the approximation of instantaneous recycling can be written
as
 \beq {{\rm d}E \over {{\rm d}t}}=\epsilon_{0}\dot
M_{\star}(700{\rm km\,s^{-1}})^{2},
 \eeq
  where $\epsilon_{0}$ is
an efficiency parameter which is still very uncertain. We take it
to be $0.02$ as in WF. The rate at which gas is heated up (to the
virial temperature) is therefore
 \beq \dot
M_{\rm heat}={0.8 \over  V_{c}^2}{{\rm d}E \over {\rm d}t}
 \eeq
which is the same form as equation (9) of Kauffmann (1996; see
also Somerville 1997). At $z=3$ and for the cosmology considered
here, this rate can be written as
 \beq\label{Mheat} \dot M_{\rm
heat} \approx 29.2h^{-0.6} \left({m_{d}\over 0.05}\right)^{1.4}
\left({\lambda \over 0.05}\right)^{-0.8}\left({V_{c}\over {250{\rm
km\,s^{-1}}}}\right)^{1.4} \left[{H(z) \over
{H_{0}}}\right]^{-0.6} {\rm M_{\odot}\,yr^{-1}}.
 \eeq
Comparing this equation with equations (7) and (8), we can find
that the rate for gas consumption due to star formation is much
larger than the rate of gas heating for LBG halos. Because LBGs
are hosted by massive halos which have large circular velocities
$V_{\rm c}$, the halos are cooling dominated which is confirmed
during the detailed calculation below. Following WF we define a
mass cooling rate by
 \beq \dot M_{\rm cool}=4\pi\rho_{\rm
gas}(r_{\rm cool})r_{\rm cool}^{2} {{\rm d}r_{\rm cool}\over {{\rm
d}t}}, \eeq
 where $r_{\rm cool}$ is the cooling radius and
$\rho_{\rm gas}$ is the density profile of the hot gas in the
halo. For simplicity, we assume that $\rho_{\rm gas}(r)=f_{\rm
B}V_c^2/(4\pi Gr^2)$, and we define $r_{\rm cool}$ to be the
radius at which the cooling time is equal to the age of the
universe, which is similar to the time interval between major
mergers of haloes (Lacey \& Cole 1994). The density distribution
of the halo mass here is assumed to be isothermal. However, it is
the NFW profile (Navarro, Frenk \& White, 1997) in MMWb. Because
the difference of the resulted cooling rates between these two
different choices of density profiles is small (Zhao et al, 1999),
and the major goal here is to show whether or not the
cooling-regulated star formation can be valid, the adoption of
isothermal profile will not influence the final result very much.

Under this definition, gas within the cooling radius can cool
effectively before the halo merges into a larger system where it
may be heated up to the new virial temperature if it is not
converted into stars. Using the cooling function given by Binney
\& Tremaine (1987) where cooling function $\Lambda \approx
10^{-23}\rm ergs^{-1}cm^{3}$ in the range of $5\times 10^{5}{\rm
K} \la T \la 2\times 10^{7}{\rm K}$ (and assuming gas with
primordial composition), the mass cooling rate can then be written
as
 \beq\label{Mcool}
  \dot M_{\rm cool} \approx  49.8
h^{1/2}\left({V_{c}\over {{\rm 250km\,s^{-1}}}}\right)^{2}
\left({f_{\rm B}\over {0.1}}\right)^{3/2} {\rm M_{\odot}{\rm
yr}^{-1}}. \eeq

If ${\dot M}_\star$ is smaller than ${\dot M}_{\rm cool}$, then
cold gas will accumulate in the halo centre and lead to higher
star formation rate. If, on the other hand, ${\dot M}_\star >{\dot
M}_{\rm cool}$, the amount of cold gas will be reduced by star
formation and supernova heating, leading to a lower star formation
rate. We therefore assume that there is a rough balance among
these three rates:
 \beq\label{balance} {\dot M_{\rm cool} \approx {\dot
M_{\rm heat} + \left(1-R_{\rm r}\right) \dot M_{\star}}}.
 \eeq
It should be noted that the cooling-regulated star formation
process is only a reasonable hypothesis, and the real situation
must be much more complicated. For example, during a major merger
of galactic haloes, the amount of gas that can cool must be much
larger than that given by the cooling argument, and the star
formation may be in a short burst (e.g. Mihos \& Hernquist 1996).
However, such bursts are not expected to dominate the observed LBG
population, because of their brief lifetimes. Thus, star formation
rates in the majority of LBGs are expected to be regulated by
equation (14) on average. As shown in MMWb, to match the observed
number density of LBGs, the median value of $V_c$ is about
$300\kms$ in the present cosmogony. The typical star formation
rate is of the order $100{\rm M_{\odot}\,yr^{-1}}$. This is not
very different from the observed star formation rates, albeit dust
distinction in the observations may be difficult to quantify.

\begin{figure}
\epsfysize=9.5cm \centerline{\epsfbox{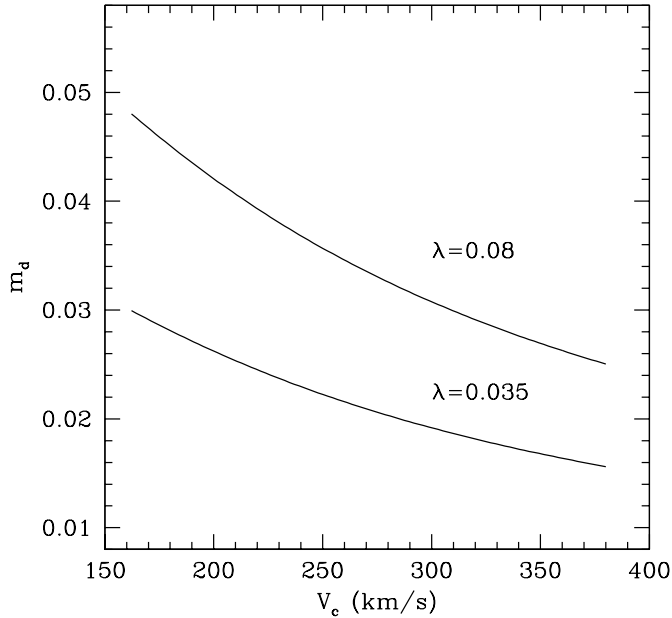}} \caption{The
value of $m_{\rm d}$ required by the balance condition equation
(14) as a function of halo circular velocity $V_{c}$ at $z=3$ for
$\lambda=0.035$ and $\lambda=0.08$, assuming $f_{\rm B}=0.1$ (see
text).}
\end{figure}

 Figure 1 shows the value of $m_{\rm d}$ required by the balance
condition equation (14) as a function of halo circular velocity,
assuming that $f_{\rm B}=0.1$ and the left hand side exactly
equals to the right hand ones in equation (14). Results are shown
for two choices of spin parameters, $\lambda=0.035$ and 0.08,
corresponding to the 50 and 90 percent points of the $\lambda$
distribution for the LBG population (MMWb). As one can see, for
the majority of LBG hosts, gas cooling indeed regulates the values
of $m_{\rm d}$ to the range from 0.02 to 0.04. So, we can
reasonably choose $m_{\rm d}=0.03$ for the LBG population as MMWb
did. Since the cooling time is approximately the age of the
universe at $z \sim 3$, cooling regulation ensures that star
formation at the predicted rate can last over a large portion of a
Hubble time.

\section {MODEL PREDICTIONS FOR THE LBG POPULATION}

Since the cooling regulation discussed above gives specific
predictions of how star formation may have proceeded in LBGs, here
we use this model to predict the properties of the LBG population.
The condition in equation (14) implies that the star formation
rate in a disk is equal to the rate of gas infall (due to a
balance between cooling and heating). Thus the evolution of the
gas in the disk of an LBG host halo is described by the standard
chemical evolution model with infall rate equal to star formation
rate, i.e., the new infalling gas to the disk distributed radially
in an exponential form with the scale length of $R_{\rm d}/b
\approx 0.7R_{\rm d}$, and the reheated gas removed decreases with
the increasing radius due to the decreasing SFR. Under the
instantaneous recycling approximation (Tinsley 1980), the gas
metallicity $Z$ is given by
 \beq\label{metallicity} Z =
y(1-e^{-\nu})+Z_{i},~~~ \nu = {\Sigma_{\rm tot}\over \Sigma _{\rm
gas}}-1,
 \eeq
where $Z_{i}$ is the initial  metallicity of the infalling gas,
$y$ is the stellar chemical yield, $\Sigma_{\rm gas}$ is the gas
surface density (which is kept constant by gas infall) and
$\Sigma_{\rm tot}$ is the total mass surface density, which
increases as star formation proceeds: \beq {{\rm d}\Sigma_{\rm
tot} \over {\rm d}t}=(1-R_{\rm r})\Sigma_{\rm SFR}. \eeq Here the
enrichment of the halo hot gases is not taken into account because
the amount of metals heated up to the halos by SNs is relatively
smaller than that of primordial gases.

\subsection{Individual Objects}

Figure 2 shows the star formation rate as a function of halo
circular velocity $V_{c}$ and spin parameter $\lambda$. As
expected, the predicted SFR increases with $V_c$ but decreases
with $\lambda$ . As we can see from the figure, if we define
systems with ${\rm SFR}\ga 40 {\rm M_{\odot}\,yr^{-1}}$ (which
matches the SFRs for the observed LBG population) to be LBGs, the
majority of their host haloes must have $V_c\ga 200\kms$ which are
cooling dominated. This result is the same as that obtained by
MMWb based on the observed number density and clustering of LBGs.
Thus, the star formation rate based on cooling argument is also
consistent with the observed number density and clustering.
Because SFR is higher in a system with smaller $\lambda$, the LBG
population are biased towards haloes with small spins, but given
its relatively narrow distribution, this bias is not very strong.

\begin{figure}
\epsfysize=9.5cm \centerline{\epsfbox{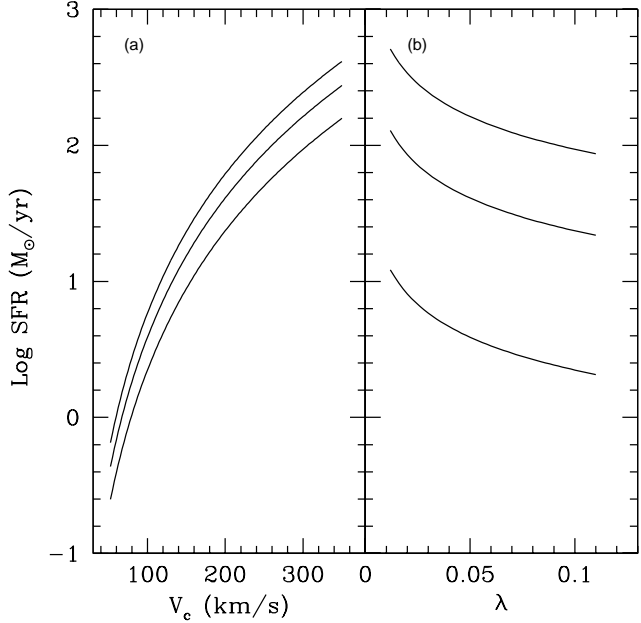}}
\caption{Predicted SFR as a function of $V_{c}$ and $\lambda$ in
the cooling-regulated model. (a) SFR vs $V_{c}$, for
$\lambda=$0.03, 0.05 and 0.1 (from top to bottom). (b) SFR vs
$\lambda$, for $V_{c}=$300, 200 and 100${\rm km/s}$ (from top to
bottom).}
\end{figure}

The predicted metallicity gradients on individual disks are shown
in Figure 3 for  two different choices of star formation time
scale $t_\star$ of 0.5Gyr and 1Gyr respectively, where we assume
that $y=Z_{\odot}$ and $Z_{\rm i}=0$ in order to make the
predictions easily compare with observations. The metallicity
gradients are negative in all cases. When radius is measured in
disk scale length, the predicted metallicity depends weakly on
$V_c$ but strongly on $\lambda$, and is higher for a longer star
formation time. As one can see from equation (15), the largest
metallicity in the model is $Z=Z_i+y$. This metallicity can be
achieved in the inner part of compact disks (with small $\lambda$)
when star formation time $t_\star\ga 1$ Gyr. The metallicity drops
by a factor of $\sim 2$ from its central value at $R\sim 3 R_{\rm
d}$.

\begin{figure}
\epsfxsize=9cm \centerline{\epsfbox{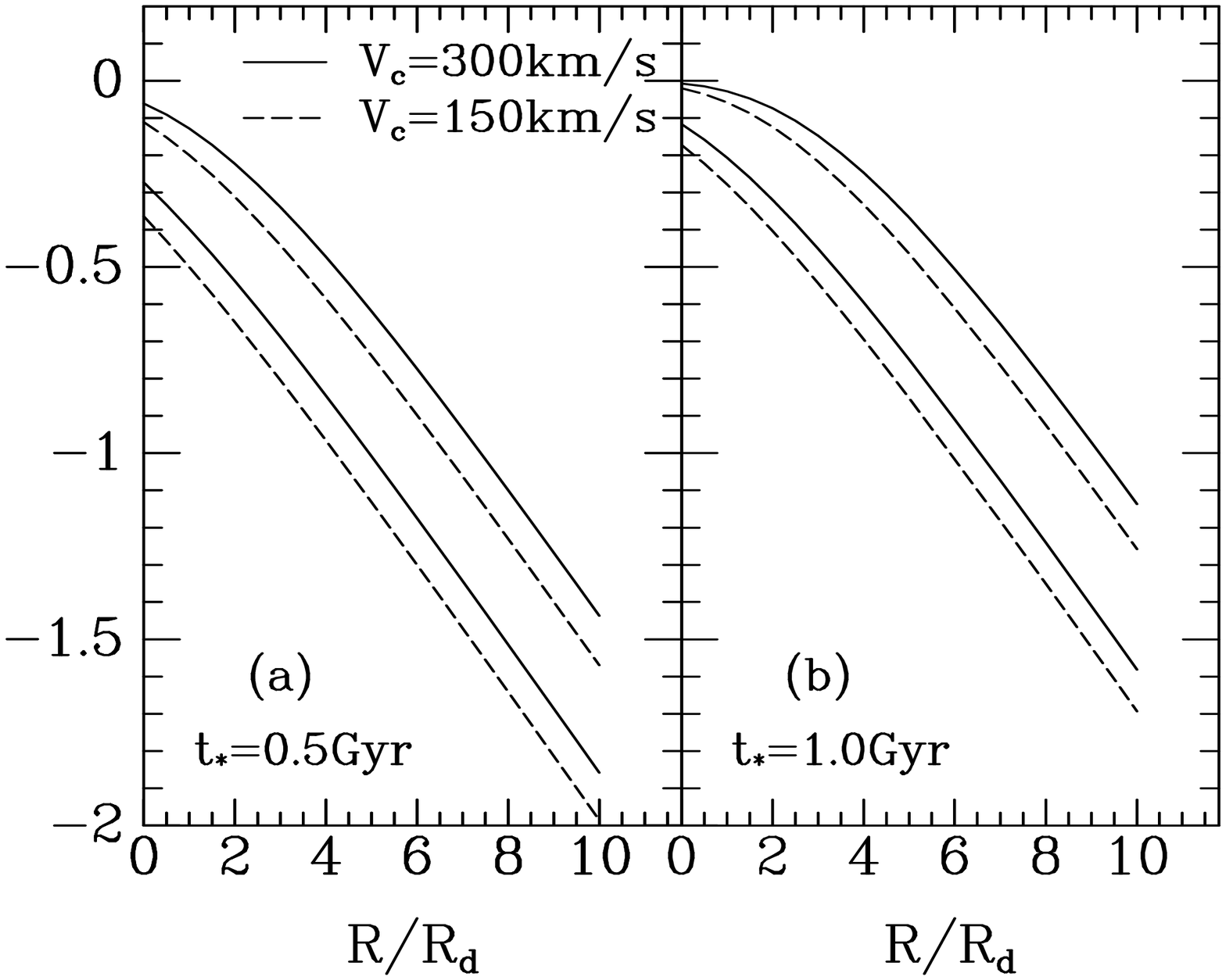}} \caption{The
metallicity gradients for LBGs for different star formation time
$t_\star$ assuming that $y=Z_{\odot}$ and $Z_{\rm i}=0$ (see
text). Full and dash lines show results for $V_{c}=300{\rm
kms^{-1}}$ and $150{\rm kms^{-1}}$, respectively. From top to
bottom, $\lambda=0.03$ and 0.1; (a) $t_\star=$0.5Gyr; (b)
$t_\star=$1Gyr}
\end{figure}

\subsection{LBG Population}

 Since  the distribution of haloes with respect to
$V_c$ and $\lambda$ are known, we can generate Monte-Carlo samples
of the halo distribution in the $V_c$-$\lambda$ plane at any given
redshift. We can then use the galaxy formation model (MMWb)
discussed above to transform the halo population into an LBG
population based on LBGs with highest SFRs which is the same as
that outlined in Sec. 2.

\begin{figure}
\epsfxsize=9.0cm \centerline{\epsfbox{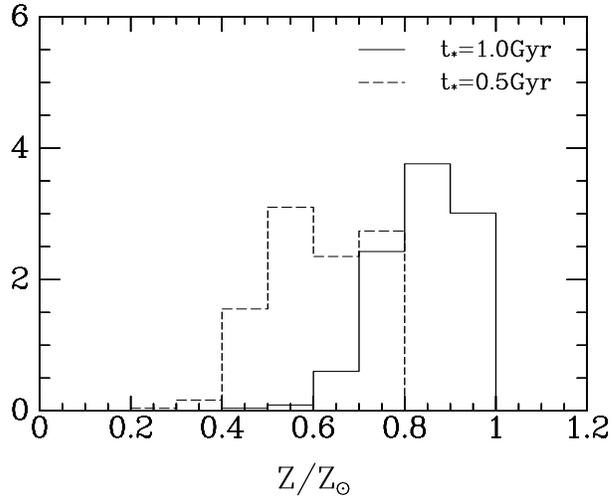}} \caption{The
predicted metallicity distributions for LBG populations assuming
that $y=Z_{\odot}$ and $Z_{\rm i}=0$ in order to make the
predictions easily compare with observations (see text). Results
are shown for two star formation timescales $t_\star=0.5$ Gyr
(dash) and $t_\star=1$ Gyr (solid), respectively (cf. equation
(15)).}
\end{figure}

We define the typical metallicity of a galaxy as the one at its
effective radius. Figure 4 shows the distribution of this
metallicity for two choices of the star formation time, $t_\star
=0.5$ Gyr and 1 Gyr. Just as the same reason as Figure 3 in last
section, we have assumed that $y=Z_{\odot}$ and $Z_{\rm i}=0$ in
order to make the predictions easily compare with observations.
The median values of $(Z-Z_{i})/y$ are 0.60 and 0.84 for
$t_\star=0.5$ Gyr and 1 Gyr, respectively. The sharp truncation at
$(Z-Z_{i})/y=1$ is due to the fact that this quantity has a
maximum value of 1 in the present chemical evolution model. It can
be inferred form Figure 3 that the range in $(Z-Z_{i})/y$
decreases with increasing star formation time. Thus, if gas infall
lasts for a long enough time, the distribution in $(Z-Z_{i})/y$
will be very narrow near 1 and all LBGs will have metallicity
$Z=Z_i+y$. According to the works of Tinsley (1980) and Maeder
(1992), the stellar yield $y$ is the order of $Z_{\odot}$ for the
Salpeter IMF. If we adopt a stellar yield $y\sim 0.5Z_{\odot}$ and
$Z_{i}=0.01Z_{\odot}$, and if LBGs are not short bursts (e.g.
$t_\star \ga 0.5$ Gyr) then their metallicity will be $Z\ga 0.2
Z_\odot$ which is similar to that proposed by Pettini (1999).

\begin{figure}
\epsfysize=9.5cm \centerline{\epsfbox{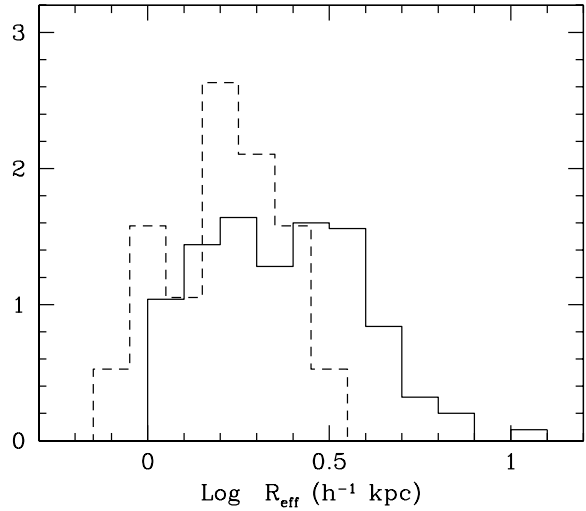}} \caption{The
predicted effective-radius distribution for LBGs in the
cooling-regulated scenario (solid), compared to the observed
distribution (dash).}
\end{figure}

The predicted distribution of effective radii for the LBG
population is shown in Figure 5.  The distribution is similar to
that of MMWb. The predicted range is $1.0\la R_{\rm eff}\la 5.0\,
h^{-1} {\rm kpc}$ with a median value of 2.5 $h^{-1}$kpc. Note
that the effective radii in the cooling-regulated model are
independent of the star formation time $t_\star$ and $m_{\rm d}$.
The model prediction is in agreement with the observational
results given by Pettini, et al. (1998), Lowenthal, et al. (1997)
and Giavalisco et al. (1996) which are mentioned above.

The predicted SFR distribution of LBGs also resembles the
prediction of MMWb except for a slight difference with MMWb, which
is shown in Figure 6. The median values are 180${\rm
M}_{\odot}~{\rm yr}^{-1}$ for the model and spans from 100 to
500${\rm M}_{\odot}~{\rm yr}^{-1}$. To compare with observations,
we have to take into account the effect of dust. If we apply an
average factor of 3 in dust extinction, then the predictions
closely match the values derived from infrared observations by
Pettini, et al. (1998) although there might exist rare LBGs with
very high SFR.

\begin{figure}
\epsfysize=9.5cm \centerline{\epsfbox{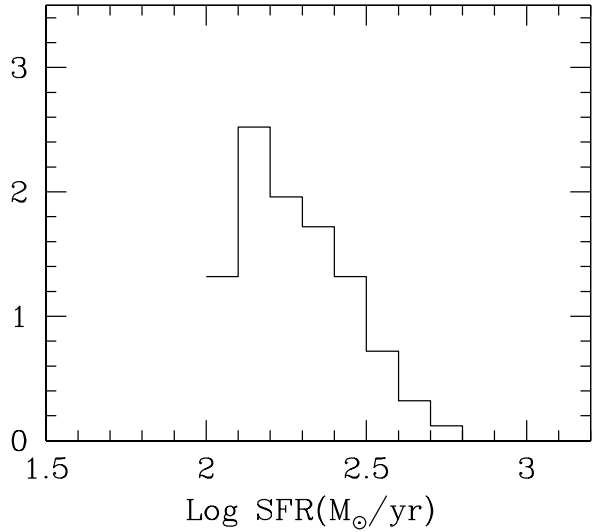}} \caption{The
predicted SFR distribution for LBGs in the cooling-regulated
scenario.}
\end{figure}

\subsection{Contribution To The Soft X-ray and UV Background}

Since the virial temperature of LBG haloes are quite high, in the
range of $10^6-10^7$K, significant soft X-ray and hard UV photons
may be emitted as the halo hot gas cools. It is therefore
interesting to examine whether the LBG population can make
substantial contribution to the soft X-ray and UV backgrounds.

The dominant cooling mechanism for hot gas with temperature $\ga
10^6$ K is the thermal bremsstrahlung. The bremsstrahlung
emissivity is given by (e.g., Peebles 1993)
 \beq
  j_{\nu}=5.4
\times 10^{-39}n_{e}^{2}T^{-1/2}e^{-h\nu / kT}{\rm erg\, cm^{-3}\,
s^{-1} \,ster^{-1}\,Hz^{-1}},
 \eeq
where $n_{e}$ (in ${\rm cm}^{-3}$) is the electron density and $T$
(in K) is the temperature given by equation (6). The total power
emitted per unit volume is
 \beq J=1.42 \times 10^{-27} ~ T^{1/2}~
n_{e}^{2} ~ {\rm erg~cm^{-3}~s^{-1}}.
 \eeq
We write the total luminosity $L_{\rm b}$ in thermal
bremsstrahlung as
 \beq L_{\rm b}=\beta {\dot M_{\rm
cool}}V_{c}^{2},
 \eeq
and we take $\beta=2.5$ here as WF so that $L_{\rm b}$ is equal to
the initial thermal energy in the cooling gas. Note that the value
of $\beta$ is quite uncertain because it depends on the detail
density and temperature profiles of the hot gas. Substituting
equation (13) into the above equation, we obtain the total soft
X-ray luminosity for an LBG
 \beq L_{\rm sx}(V_c)\approx 4.1 \times
10^{40} f_{\rm soft} \left({V_{c} \over {\rm
250km/s}}\right)^{4}\left({f_{\rm B} \over 0.1}\right)^{3/2} {\rm
erg ~ s^{-1}}, \eeq where \beq f_{\rm soft} = {1 \over kT}
{\int_{0.5(1+z)}^{2(1+z)}} e^{-E/kT} dE
 \eeq
is the fraction of total energy that falls into the ROSAT soft
X-ray (0.5-2 keV) band. The contribution of the LBG population to
the soft X-ray background is then
 \beq \rho_{\rm sx} = \int\int
dV_c dV_{\rm com} { n(z) L_{\rm sx}  \over 4\pi d_{L}^2} \approx
5.7 \times10^{-8} \left({f_{\rm B} \over 0.1}\right)^{3/2} {\rm
erg ~s^{-1}cm^{-2}},
 \eeq
where $n(z)$ is the comoving number density of LBG haloes as a
function of redshift $z$, $dV_{\rm com}$ is the differential
comoving volume from $z$ to $z+dz$ and $d_{L}$ is the luminosity
distance. The integrate for $V_{\rm c}$ is to sum up all selected
LBGs with $V_{\rm c}$ based on their highest SFRs. We have
integrated over redshift range from 3 to 4 where the number
density of LBGs is nearly a constant (Steidel 1998a,b). This
contribution should be compared with the value derived from the
ROSAT observations (Hasinger et al. 1998) in the 0.5-2 keV band
\beq \rho_{\rm sx}\approx 2.4\times10^{-7}{\rm
erg\,s^{-1}cm^{-2}}. \eeq As we can see, the soft X-ray
contribution from LBGs could be a substantial fraction (about
20\%) of the total soft X-ray background.

Similarly we can calculate the contribution of LBGs to the UV
background at $z=3$. We evaluate the UV background at 4 Ryd
(1Ryd=13.6 eV) using nearly identical procedures, we find that
\beq i_{\rm 4Ryd} \approx 2.4\times10^{-24}{ \left({f_{\rm B}
\over 0.1}\right)^{3/2} {\rm ergs^{-1}cm^{-2}Hz^{-1}ster^{-1}}},
\eeq which is much smaller than the UV background from AGNs,
$i_{\rm 4Ryd} \sim 10^{-22}{\rm ergs^{-1}cm^{-2}Hz^{-1}ster^{-1}}$
(e.g. Miralda-Escude \& Ostriker 1990).

\subsection{Contribution to the Total Metals}

Based on the recent observational results of the cosmic star
formation history, Pettini (1999) obtained a predicted total mass
of metals produced at $z=2.5$. After combining results of all
contributors observed, he argued that there seems to exist a very
serious ``missing metal" problem, i.e., the predicted result is
much higher than observed ones. So, it is interesting to evaluate
the total metals produced by LBGs in our model.

According to the method we select LBGs to be the galaxies with
highest SFR and our chemical evolution model mentioned in Sec.
3.2,  we can calculate the total metal density produced by the LBG
population at $z=3$ based on their observed comoving number
density which is $N_{\rm LBG}=2.4 \times 10^{-3}h^{3}{\rm
Mpc^{-3}}$ for the assumed cosmology (Adelberger et al. 1998).
Defining that $\Omega_{\rm Z}$ is the metal density relative to
the critical density, we get that $\Omega_{\rm Z}$ of LBGs are
$0.19\Omega_{\rm B}\times y$ and $0.29\Omega_{\rm B}\times y$ for
star formation time of 0.5Gyr and 1Gyr respectively, where $y$ is
the stellar yield which is the same as above. Because the virial
temperature of LBG halos are very high, a significant fraction of
the metal should be in hot phase. Comparing our results with that
estimated by Pettini (1999) which is $0.08\Omega_{\rm B}\times
Z_{\odot}$ (the cosmogony has been taken into account), we find
that there is no ``missing metal" problem in our model.

\subsection{LBGs and Damped Lyman-Alpha Systems}

Damped Lyman-alpha systems (DLSs) are another population of
objects that can be observed at similar redshift to LBGs. The DLSs
are selected according to their high neutral HI column density
($>10^{20.3}~{\rm cm}^{-2}$), and  are believed to be either
high-redshift thick disk galaxies (Prochaska \& Wolfe 1998) or
merging protogalactic clumps (Haehnelt, Steimetz \& Rauch 1998).
In either case, to match the observed abundance of DLSs, most DLSs
should have circular velocity between $50\kms$ to $200\kms$, much
smaller than the median circular velocity of LBGs ($\sim
300\kms$). Based on the PS formalism (equation (5)) and disk
galaxy formation scenario suggested by MMWa (equations (1) and
(2)), we can estimate with the random inclination being taken into
account, that the fraction of absorbing cross-sections contributed
by LBGs amounts to only about 5\% of the total absorption
cross-section assumed LBGs with highest SFRs. This means that only
a very small fraction of DLSs can be identified as LBGs.

The physical connection between LBGs and DLSs is still unclear,
although the recent observation of Moller \& Warren (1998) using
$HST$ indicates that some DLSs could be associated with LBGs. In
Figure 7, we show the predicted metallicity distribution for the
subset of DLSs which can be observed as LBGs. aGain, we have
assumed that $y=Z_{\odot}$ and $Z_{\rm i}=0$ to let the
predictions more easily compare to observations. As can be seen,
the DLSs generally have lower metallicity than LBGs, because they
are biased towards the outer region of the host galaxies, where
the star formation activity is reduced. Notice, however, that the
metallicity of these DLSs could still be higher than most DLSs at
the same redshift, which typically have metallicity of $0.1
Z_\odot$ (Pettini, et al 1997a).

\begin{figure}
\epsfxsize=9.cm \centerline{\epsfbox{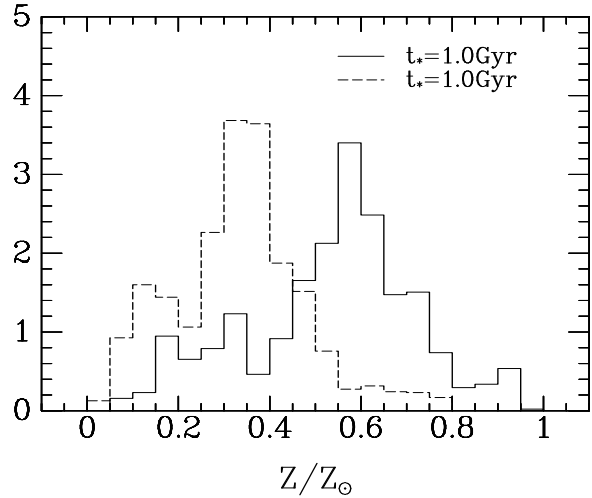}} \caption{The
predicted metallicity distribution for the DLSs expected from the
LBG population (see text). Results are shown for two star
formation timescales $t_\star=0.5$ Gyr (dash) and $t_\star=1$ Gyr
(solid), respectively.}
\end{figure}

\section{SUMMARY}

In this paper, we have examined the star formation and chemical
enrichment in Lyman break galaxies, assuming them to be the
central galaxies of massive haloes at $z\sim 3$ and using simple
chemical evolution models. We found that gas cooling in dark
haloes provides a natural process which regulates the amount of
star forming gas. The predicted star formation rates and effective
radii are consistent with observations. The metallicity of the gas
associated with an LBG is roughly equal to the chemical yield, or
about the order of $1 Z_{\odot}$ for a Salpeter IMF. Because of
the relatively long star-formation time, the colours of these
galaxies should be redder than that of short starbursts. It is not
clear whether this prediction is consistent with current (rather)
limited observations, because the interpretation of the
observational data depends strongly on the adopted dust reddening.
Stringent constraint can be obtained when full spectral
information of the LBG population is carefully analyzed.

The model predicts a marked radial metallicity gradient in an LBG,
with the gas in the outer region having lower metallicity. As a
result, the metallicities for the damped Lyman-alpha absorption
systems expected from the LBG population are lower than those for
the LBGs themselves, although high metallicity is expected for a
small number of sightlines going through the central regions of an
LBG. At the same time, our modeled contribution to the total metal
is roughly consistent with that obtained from the observed cosmic
star formation history, i.e., there might not exist so-called
``missing metal" problem although there could be more than half of
the metals to be in the hot phase. Finally, a prediction of our
model is that LBG haloes are filled with hot gas. As a result,
these galaxies may have a non-negligible contribution to the soft
X-ray background. The contribution of LBGs to the ionizing UV
background is found to be small.

There are two basic assumptions in our work. One is that the LBG
population is one-to-one associated with the most massive halos
which are generated from the PS formalism, as done by MMWb;
another is that the timescale of star formation for LBG population
is assumed to be the order of 1Gyr, which is suggested by Steidel
et al. (1999a,b, 1995). However, Baugh et al. (1999) recently
argue that the prediction of the clustering properties of LBGs
based on this first simple assumption will be discrepancy with the
results of more detailed semi-analytic models.  Still, the second
will lead to difficulty in reproducing the redshift evolution of
bright galaxies (Kolatt et al. 1999). More detailed modelling done
by Somerville (1997) suggest the collisional starbursts could be
expected to be an important effect in understanding the LBGs. So,
further observations are required to investigate the intrinsic
properties of LBGs.

\section*{Acknowledgments}

This project is partly supported by the Chinese National Natural
Foundation. I thank Dr. S. Mao, Dr. H. J. Mo and Prof. S. D. M.
White for detailed discussions, and the useful help of anonymous
referee.

{}

\begin{thebibliography}{}
\bibitem{} Adelberger,K.L., Steidel,C., $\&$ Giavalisco,M., et al., 1998,
ApJ, 505, 18
 \bibitem{} Bardeen, J.M., Bond,
J.R.,$\&$ Kaiser, N., et al., 1986, ApJ, 304, 15
\bibitem{} Baugh,C.M., Cole,S.,$\&$ Frenk,C.S.,1998,
preprint(astro-ph/9808209)
\bibitem{} Baugh,C.M., Benson, \& Cole,S., et al., 1999, MNRAS, 305, L21
\bibitem{} Binney,J. $\&$ Tremaine,S., 1987,
Galactic dynamics. Princeton Univ. Press, Princeton, NJ, P580
\bibitem{}Burles,S., \& Tytler,D., 1998, ApJ, 507, 732
\bibitem{}Calzetti,D., 1997, AJ, 113, 162
\bibitem{} Coles,P., Lucchin,F.,
$\&$ Matarrese,S., 1998, MNRAS, 300, 183
\bibitem{} Cole,M. $\&$ Lacey,C., 1996,
preprint(astro-ph/9510147 v3)
\bibitem{} Friaca,A.C.S. \&
Terlevich,R.J., 1999, MNRAS, 305, 90
\bibitem{}
Giavalisco,M., Steidel,C., $\&$ Adelberger,K.L., 1998, ApJ, 503,
543
\bibitem{} Giavalisco,M., Steidel,C.,
$\&$ Macchetto,F.D., 1996, ApJ, 470, 189
\bibitem{}Governato,F., Baugh,C.M., $\&$ Frenk,C.S., et al., 1998,
Nature, 392, 359
\bibitem{} Haehnelt M., Steinmetz M., Rauch
M.,1998, ApJ, 495, 647
\bibitem{} Hasinger,G., Burg,R., $\&$
Giacconi,R., et al., 1993, A$\&$A, 275, 1
\bibitem{} Jing,Y.P.,
1998, ApJ, 503, L9
\bibitem{} Jing,Y.P. $\&$
Suto,Y., 1998, 494, L5
\bibitem{}Kauffmann,G., 1996, MNRAS, 281, 475
\bibitem{}Kauffmann,G., Colberg,J.M., Diaferio,A., $\&$ White,
S.D.M., 1999, MNRAS, 303, 188
\bibitem{} Katz,N.,
Hernquist,L.,  $\&$ Weinberg,D.H., et al., 1998,
preprint(astro-ph/9806257)
\bibitem{} Kennicutt,R., 1998, ApJ, 498, 541
\bibitem{} Kolatt, et al., 1999, preprint (astro-ph/9906104)
\bibitem{} Lacey,C. $\&$ Cole,S.,
1994, MNRAS, 271, 676
\bibitem{}Lowenthal,J.D., Koo,D.C., $\&$
Guzman,R., et al, 1997, ApJ, 481, 673
\bibitem{} Madau,P.,
Pozzetti,L., $\&$ Dickinson,M., 1998, ApJ, 499, 106
\bibitem{}
Madau,P., Ferguson,H.C., $\&$ Dickinson,M., et al., 1996, MNRAS,
283, 1388
\bibitem{} Maeder,A., 1992, A$\&$A, 264, 105
\bibitem{} Mihos,J.C. $\&$ Hernquist,L., 1996, ApJ, 464, 641
\bibitem{} Miralda-Escude,J. \& Ostriker.J.P., 1990, ApJ, 350, 1
\bibitem{} Mo, H.J.  $\&$ Fugugita,M., 1996, ApJ, 467, L9
\bibitem{} Mo, H.J., Mao, S., $\&$ White, S.D.M., 1998a, MNRAS,
295, 319(MMWa)
\bibitem{} Mo, H.J., Mao, S., $\&$ White, S.D.M.,
1998b, 1999, MNRAS, 304, 175 (MMWb)
\bibitem{} Moller,P.  $\&$ Warren,S.J., 1998, MNRAS, 299, 661
\bibitem{} Moscardini,L., Coles,P., $\&$  Lucchin,F., et al.,
1998, MNRAS, 299, 95
\bibitem{} Narvarro, J. F., Frenk, C. S., \& White, S. D. M.,
1997, ApJ, 490,493
\bibitem{} Ouchi,M. \&
Yamada,T., 1999, ApJ, 517, L19
\bibitem{} Peacock,J.A., Jimenez,R.,  $\&$ Dunlop,J.S., et al., 1998,
preprint (astro-ph/9801184)
\bibitem{} Peebles,P.J.E., 1993,
Principles of Physical Cosmology, Princeton Univ. Press,
Princeton, NJ, P577
\bibitem{} Pettini,M., 1999, preprint
(astro-ph/9902173)
\bibitem{} Pettini,M., Smith,L.J., King,D.L.,
$\&$ Hunstead,R.W., 1997a, ApJ, 486, 665
\bibitem{} Pettini,M.,
Steidel,C., Dickinson M., Kellogg,M., Giavolisco M., $\&$
Adelberger K.L., 1997b, preprint (astro-ph/9707200)
\bibitem{}
Pettini,M., Kellogg,M., $\&$ Steidel,C., et al., 1998, ApJ, 508,
539
\bibitem{}Press,W.H. $\&$ Schechter,P., 1974,
ApJ, 187, 425 (PS)
\bibitem{} Prochaska,J.X. \& Wolfe,A.M.,
1998, ApJ, 507, 113
\bibitem{} Sawicki M., Yee
H.K.C., 1998, AJ, 115, 1329
\bibitem{} Somerville,R.S.,  1997,
PhD Thesis
\bibitem{} Somerville,R.S.,  Primack,J.R., \&
Faber,S.M.,1999, MNRAS, 307, 15
\bibitem{}
Steidel,C., Pettini,M., $\&$ Hamilton,D., 1995, AJ, 110, 2519
\bibitem{}Steidel,C., Giavalisco,M., $\&$ Pettini,M., et al.,
1996, ApJ,462, L17
\bibitem{} Steidel,C., Adelberger,K.L., $\&$
Dickison,M., et al., 1998, ApJ, 492, 428
\bibitem{} Steidel,C.,
Adelberger,K.L., $\&$ Giavalisco,M., et al., 1999a, ApJ, 519, 1
\bibitem{} Steidel,C., Adelberger,K.L., $\&$
Dickison,M., et al., 1999b, preprint, (astro-ph/9812167)
\bibitem{} Tinsley,B.M., 1980, Fundam. Cosmic Phys., 5, 287
\bibitem{} Toomre,A., 1964, ApJ, 139, 1217
\bibitem{}
Wechsler,R.H.,Gross,M.A.K.,  $\&$ Primack,J.R., et al., 1998, ApJ,
509, 19
\bibitem{} White,S.D.M. $\&$
Frenk,C.S., 1991, ApJ, 379, 25 (WF)
\bibitem{} White,S.D.M. $\&$
Rees,M.J., 1978, MNRAS, 183, 341
\bibitem{} Zhao, D., Shu, C., Song, G., \& Zhao, J., 1999,
submitted to ApJ
\end{thebibliography}
\end{document}